\documentclass[acmsmall]{acmart}
\usepackage{multirow}
\usepackage[normalem]{ulem} 
\usepackage{tabularx}
\usepackage{xcolor}

\definecolor{Issue1}{HTML}{ff7e16}
\definecolor{Issue2}{HTML}{2476b3}
\definecolor{Issue3}{HTML}{309f30}
\definecolor{Issue4}{HTML}{d52b2c}

\def\BibTeX{{\rm B\kern-.05em{\sc i\kern-.025em b}\kern-.08emT\kern-.1667em\lower.7ex\hbox{E}\kern-.125emX}}
    
\copyrightyear{2020}
\acmYear{2020}
\setcopyright{acmlicensed}
\acmConference[CSCW '20]{CSCW '20: ACM Conference on Computer Supported Collaborative Work and Social Computing}{October 17--21, 2020}{Minneapolis, MN}
\acmBooktitle{CSCW '20: ACM Conference on Computer Supported Collaborative Work and Social Computing, October 17--21, 2020, Minneapolis, MN}

\setcopyright{acmcopyright}
\acmJournal{PACMHCI}
\acmYear{2020} \acmVolume{4} \acmNumber{CSCW1} \acmArticle{68} \acmMonth{5} \acmPrice{15.00}\acmDOI{10.1145/3392878}

\received{January 2020}
\received[accepted]{March 2020}

\begin{document}

%
\title[Human Factors in Model Interpretability: Industry Practices, Challenges, and Needs]{Human Factors in Model Interpretability: Industry Practices, Challenges, and Needs}

%
\author{Sungsoo Ray Hong}
\email{rayhong@nyu.edu}
\affiliation{
  \institution{New York University}
  \city{New York City}
  \state{NY}
  \country{USA}
}

\author{Jessica Hullman}
\email{jhullman@northwestern.edu}
\affiliation{
  \institution{Northwestern University}
  \city{Evanston}
  \state{IL}
  \country{USA}
}

\author{Enrico Bertini}
\email{enrico.bertini@nyu.edu}
\affiliation{
  \institution{New York University}
  \city{New York City}
  \state{NY}
  \country{USA}
}

%
\renewcommand{\shortauthors}{Sungsoo Ray Hong et al.}

%
\begin{abstract}
As the use of machine learning (ML) models in product development and data-driven decision-making processes became pervasive in many domains, people's focus on building a well-performing model has increasingly shifted to understanding \textit{how} their model works. While scholarly interest in model interpretability has grown rapidly in research communities like HCI, ML, and beyond, little is known about how practitioners perceive and aim to provide interpretability in the context of their existing workflows. This lack of understanding of interpretability as practiced may prevent interpretability research from addressing important needs, or lead to unrealistic solutions. To bridge this gap, we conducted 22 semi-structured interviews with industry practitioners to understand how they conceive of and design for interpretability while they plan, build, and use their models. Based on a qualitative analysis of our results, we differentiate interpretability roles, processes, goals and strategies as they exist within organizations making heavy use of ML models. The characterization of interpretability work that emerges from our analysis suggests that model interpretability frequently involves cooperation and mental model comparison between people in different roles, often aimed at building trust not only between people and models but also between people within the organization. We present implications for design that discuss gaps between the interpretability challenges that practitioners face in their practice and approaches proposed in the literature, highlighting possible research directions that can better address real-world needs.
\end{abstract}

%
%

%

%

%
\maketitle

\section{Introduction}

Recent years have witnessed a rapid increase in the deployment of machine learning (ML) in a large variety of practical application areas, such as finance~\cite{ceena2018towards}, healthcare~\cite{balagopalan2018effect, esteva2017dermatologist, krause2016interacting}, governance~\cite{meijer2019predictive}, education~\cite{holstein2018student}, space exploration~\cite{Mjolsness2051}, and digital product development~\cite{he2014practical,mcmahan2013ad}. This rapid increase in adoption comes with an increased concern about humans' capabilities to understand how ML models work and to design models that are guaranteed to behave as expected in critical circumstances. Motivating this concern are the fact that models are increasingly used for sensitive applications, where mistakes can cause catastrophic consequences for organizations and individuals~\cite{luo2016automatically}, and that ML technologies have grown exponentially in complexity (e.g., with the advent of deep learning), making it hard to verify how models ``think'' and to predict their future behavior. Growing recognition of the need to understand ``human factors'' in ML model development and use is reflected in the recent shift from a unidirectional focus on model accuracy, to a larger perspective that includes a strong focus on machine learning \textit{interpretability}~\cite{molnar2019Interpretable, pandey_2019}.

In addition to proposing new algorithms and tools~\cite{breiman2017classification, letham2015interpretable, ribeiro2016should, simonyan2013deep, wang2017bayesian}, the growing literature on interpretability includes attempts to define interpretability with scientific rigor~\cite{doshi2017towards, doshi-Velez2018consideration,kim2016examples,schmidt2019quantifying}. Interpretability is generally intended to arise from an alignment between the human user's mental model of how a system will behave and the system's actual behavior. Achieving this alignment can lead to measurable benefits such as a better performance of a human-model team on a task~\cite{bansal2019} or the user's improved ability to debug the model~\cite{kulesza2015principles}. Therefore, communities in CSCW and Human-Computer Interaction (HCI) have devoted effort to understanding which ML models can better conform to human understanding of prediction tasks~\cite{zhang2019dissonance}, how people's interaction with ML models affects the quality of decisions~\cite{green2019principles}, and how data scientists use particular tools in conducting their interpretability tasks~\cite{kaur2020interpreting}.


However, despite this rapid expansion in tools and formal and empirical definitions, our understanding of how interpretability is understood by and how it impacts the actual ML practitioners who work ``in the trenches'' remains elusive. While some discussion of practitioners' needs is often included in research works that propose novel systems~\cite{ren2016squares, talbot2009ensemblematrix,wongsuphasawat2017visualizing,zhang2019manifold}, there has been a relative lack of studies aimed at empirically understanding how ML professionals perform interpretability-related tasks and what are their practices, needs, and challenges. For instance, while interpretability is often framed as how well a model communicates its decisions to a user, much less is known about how interpretability manifests in real workplaces in which teams must communicate and coordinate their work around models and decision-making tools. We believe bridging this gap is necessary because it can help researchers ground their work in the real problems practitioners face when dealing with interpretability issues.

In this work, we contribute with an empirical study investigating how model interpretability is practiced, perceived, and negotiated by data scientists and other professionals with different roles in industry. We conducted semi-structured interviews with 22 ML practitioners from 20 different companies who are responsible for building models that are state-of-the-art in their field. Some of the models they built are used by millions of daily active users and/or support critical decisions where life is at stake. Our interviewing approach aimed to differentiate how our interviewees perceived interpretability \textit{roles}, \textit{processes}, \textit{goals}, and associated challenges.

Our results contribute novel insights into the three roles (i.e., model \textit{builders}, model \textit{breakers}, and model \textit{consumers}) played by different stakeholders involved in interpretability work; how their tasks and strategies vary depending on three stages in their model building process (i.e., model \textit{conceptualization}, model \textit{building}, and model \textit{deployment}) in order for practitioners to achieve model interpretability and which types of roadblocks they often encounter. Importantly, the characterization that emerges from our results sheds light on the underexplored social nature of interpretability as it manifests in ML efforts in an organization. Our results suggest that ML interpretability is contextually dependent in terms of practitioners' role within their organization and socially negotiated through interactions between different stakeholders at different stages in the development pipeline. Practitioners appear to perceive model interpretability as more than simply aligning an ML model with a single notion of the human mental model to achieve interpretability as a property of the model: it is simultaneously about the process of negotiating interpretability and its role in creating insight and decisions that occurs through considering the mental models between different stakeholders. While interpretability work is intended to foster trust in models, as discussed in existing interpretability literature, our interviewees also highlighted how interpretability plays an equally important role in building organizational trust among data scientists and other stakeholders through the three aforementioned stages of conceptualization, building, and deployment. 

These results suggest important avenues for future work aimed at bridging the gap between interpretability as understood in research versus practice. We conclude by using our results to motivate design opportunities for technology aimed at addressing core communication-oriented challenges of interpretability work in organizations. 

\section{Related Work}

We discuss the motivations behind model interpretability research and present key approaches to understanding and realizing interpretability in HCI, CSCW, and ML communities. We explain the limitations and criticism discussed in the literature. Finally, we describe relevant empirical studies that motivated our work.

\subsection{ML Interpretability: Background and Approaches}

Models are often categorized into two classes in light of the ease with which they can be interpreted by humans: \textit{white-box} models and \textit{black-box} models. White-box models tend to have a transparent structure that permits human observers to understand the logic that drives model decisions. Examples include decision trees, logistic regression, and many varieties of rules. Black-box models have complex structures that create much less intelligible relationships between input and output. Examples include a large variety of neural networks, random forests, and ensemble models.

The fast-growing interest in model interpretability is closely related to the raise of black-box models as practitioners are concerned with their opacity. But white-box models can also be troublesome as their structure, though transparent, can become very large and complex. For instance, linear models with a high number of variables, trees with a high number of nodes and rule lists with a large set of conditions also pose interpretability challenges to humans.

While successful machine-human collaboration can yield outcomes that outperform either human or machine acting alone~\cite{bansal2019, kneusel2017improving, patel2011using, choi2019AILA}, how to align human mental models with ML models in ways that are helpful rather than harmful has become crucial in practice, and remains a topic of active research. For example, even though black-box models are known to provide better performance than white-box models in many cases~\cite{guidotti2019survey}, over-trusting models may result in an increased risk of making a decision with severe negative consequences~\cite{ceena2018towards} and/or lower trust in model predictions~\cite{luo2016automatically}.
As several studies have shown, ML models can learn spurious associations~\cite{gilpin2018explaining, caruana2015intelligible} and make decisions based on correlations that do not result from causal relationships. Models can also learn relationships in data that exemplify undesirable forms of bias~\cite{kamiran2009classifying} or are mislabeled~\cite{chestxray14}. Hence in practice, model accuracy cannot be fully trusted as a reliable metric for automation and decision making~\cite{krause2016interacting}, grounding a need for interpretability efforts.

Model interpretability techniques are often categorized into two classes: \textit{global} and \textit{local}~\cite{doshi-Velez2018consideration, guidotti2019survey, molnar2019Interpretable}. \textit{Global interpretability} refers to showing the logical structure of a model to explain how it works globally. This type of information is readily available in white-box methods with the caveat mentioned above that even transparent models can become hard to understand when they become too large~\cite{ming2018rulematrix}. When the goal is to understand a black-box model at a global level \textit{model extraction} methods are needed, that is, methods that use the input-output behavior of the model to infer a structure able to describe its behavior. Examples include methods that infer feature relevance from the model~\cite{friedman2001greedy} as well as methods that build whole surrogates that try to mimic the model's behavior by building a more transparent equivalent~\cite{craven1996extracting}.

\textit{Local interpretability} is about providing explanations at a single instance level. Some approaches, such as LIME~\cite{ribeiro2016should} and SHAP~\cite{Lundberg2017SHAP} focus on providing a set of ``weights'' aimed at providing insights about how an instance is processed by the model. Anchors aims at providing instance explanations using a rule structure~\cite{ribeiro2018anchors}. Some methods provide explanations more in the form of ``counterfactual'', that is, the minimal set of changes necessary to change the instance's predicted outcome~\cite{martens2014explaining}.

In addition to model extraction techniques, researchers have developed interactive visualization interfaces to help data scientists' build, interpret, and debug models~\cite{hohman2019gamut}. For instance, EnsembleMatrix presents multiple confusion matrices to help combine multiple models~\cite{talbot2009ensemblematrix}. Squares visualizes the performance of multiple models in a single view, enabling fast model comparison~\cite{ren2016squares}. Some other approaches present visual analytics systems specialized in a specific model type, such as regression models~\cite{dingen2019regressionexplorer}, neural networks~\cite{wongsuphasawat2017visualizing, Kahng2017activis, smilkov2017direct}, and surrogate modes~\cite{ming2018rulematrix}.

\subsection{Limitations and Criticism of Model Interpretability}

As interpretability research grows, critiques have also grown more prominent. While the existing literature emphasizes the importance of model interpretability as a tool that humans can leverage to align their mental model with ML models~\cite{kulesza2015principles}, some studies paint a less positive picture.

For example, Bussone et al., found users of clinical decision-support models tend to over-rely on a model recommendation over their own expertise~\cite{bussone2015role}. Aligning with their findings, Kaur et al., found that data scientists often ``misuse'' interpretability tools and over-trust the results~\cite{kaur2020interpreting}. Narayanan et al., investigate how explanation complexity (defined as explanation size, cognitive chunks, and the number of repeated terms) impacts accuracy, efficiency and user satisfaction. Initial results show that explanation complexity has a negative impact on efficiency and satisfaction but not necessarily on accuracy~\cite{menaka2018how}. Green and Chen's findings suggest that people's decisions when presented with model predictions may not achieve desiderata related to reliability and fairness~\cite{green2019principles}. Forough et al., investigated how the number of input features and model transparency affect simulatability (the user's ability to predict model behavior) and trust~\cite{forough2018manipulating}. Their results suggest that simpler and more transparent models don't necessarily help people build better mental models and that transparency may even hinder people's ability to correct inaccurate predictions. This line of research suggests that the impact of model interpretability can be complex, and may be dependent on properties of the consumers and context on the top of the tools available in practice.

More broadly, scholars have argued that the concept of model interpretability is ill-defined and couched in ambiguous terminology~\cite{bibal2016interpretability, Lipton16a, forough2018manipulating}. For example, Lipton discusses how although one of the goals for interpreting models is gaining trust, the intended meaning of trust can vary, from faith in a model's ability to perform accurately, to acquiring a ``low-level mechanistic understanding'' of a model~\cite{Lipton16a}. The use of many different terms, such as comprehensibility, mental fit, explainability, acceptability, and justifiability, to refer to similar concepts ~\cite{bibal2016interpretability} makes it hard to understand interpretability's scope and to define metrics for its evaluation~\cite{forough2018manipulating}.

Such critiques have prompted reflection and at times more formal definitions. Doshi-Velez and Kim see model interpretability as a proxy for many ``auxiliary criteria'' such as safety, fairness, non-discrimination, avoiding technical debt, privacy, reliability, providing the right to explanation, trust, and more~\cite{doshi2017towards}. 
Attempts to operationalize interpretability approximate it using the relative time it takes a human or group of humans (such as crowd workers) to predict a model's label for an instance~\cite{lage2018human}, or the overlap between human annotations for multiple instances and those generated by a model in text classification tasks~\cite{schmidt2019quantifying} and image recognition tasks~\cite{zhang2019dissonance}.

Recent studies suggest that the contexts in which data scientists work with ML models are diversifying~\cite{zhang2020data}. A deeper understanding of how practitioners work with model interpretability across contexts in the real world is necessary for developing a science of interpretability~\cite{kaur2020interpreting}. For instance, Miller argues that the majority of studies discuss model interpretability based on the unit of an individual, but few studies explain how humans describe how models work to one another~\cite{miller_2019}. While some recent work presents insights regarding how laypeople use automated tools in building ML models~\cite{Yang2018Grounding, wang2020humanAI} or ML experts use specific data science packages in determining insights related to model interpretability~\cite{kaur2020interpreting}, little research to date focuses on investigating holistic model interpretability processes or communication strategies among different roles in organizations. Inspired by studies that emphasize workplace settings~\cite{luff2000workplace, zhang2020data}, our work characterizes interpretability via how data scientists and others involved in this work perceive the different stakeholders and strategies that characterize interpretability work in organizations using ML. In doing so, our work problematizes attempts to define interpretability through measurable, fixed aspects of a single human's interactions with a model by uncovering how those engaged with interpretability portray it as an ongoing result of negotiations between people within an organization.

Our ultimate goal is to help the scientific community ground their work on practical needs by describing current practices and gaps that need to be bridged. Our work is therefore in line with similar attempts at providing a window to the world of practitioners in ML and Data Science, such as Holstein et al.'s interview study on ML fairness~\cite{Holstein2019improving}; Yang et al.'s, survey on how laypeople use ML~\cite{Yang2018Grounding}; and Kandel et al.'s interview study on how enterprise analysts use data analysis and visualization tools~\cite{kandel2012enterprise}.

\section{Research Method}

The goals of our research are to (1) characterize how ML experts conceive of interpretability and how interpretability manifests in their organization, and (2) identify aspects of interpretability practices that are not sufficiently supported by existing technology. To achieve these goals, we conducted open-ended conversational interviews with professional ML experts. While ethnographic methods like participant observation are also likely to support these goals, based on the lack of prior work focusing on interpretability work in real organizational settings, we opted for interviews as they allowed us to reach a broader set of people working with ML in industry, helping ensure our results were general enough to apply across multiple domains. We describe the methodology we followed to conduct the interviews and to analyze the data we collected.

\subsection{Recruiting Participants}
To recruit practitioners, we used convenience and snowball sampling strategies~\cite{creswell2016}. We first communicated with industry acquaintances who are building and using state-of-the-art ML solutions in their fields. We invited these contacts to participate in a voluntary interview if they had had personal experience with model interpretability issues and solutions, and/or needed to communicate the results of their ML models to others. We also asked them to suggest practitioners from other companies and domains who may have relevant experience for our study. This snowball sampling helped ensure that we recruited participants representing a variety of application areas and enterprise settings. As we continued our interviews, we identified new directions in need of further investigation as our understanding of several initial topics became saturated.

\begin{table}
  \centering
  \begin{tabularx}{\columnwidth}{l l l X}
    \toprule
    {\textit{PID}}
    & {\textit{Company domain}}
      & {\textit{Job title (role)}}
    & {\textit{Domain problems}} \\
    \midrule
    P1 & Software & Staff manager in Research (DS) & Object detection for telepresence robots \\
    P2 & Consulting & CEO (DS) & Identity recognition, fraud detection \\
    P3 & Banking & Senior ML Engineer (DS) & Credit risk assessment, call transcription \\
    P4 & Software & Lead Data Scientist (DS) & Sentiment analysis, object detection, AutoML \\
    P5 & Banking & Data Engineer (SE) & Anomalous recurring online payment detection \\
    P6 & Banking/Finance & Head of AI (DS) & Credit evaluation for business loans \\
    P7 & Internet Services & Senior Software Developer (SE) & Tooling for tabular/image data \\
    P8 & Social Media & Data Scientist (DS) & Service user retention prediction \\
    P9 & Banking/Finance & Principal Data Scientist & Customer acquisition/financial forcasting \\
    P10 & Software & Product Manager (PM) & Tooling (visualizing model interpretation) \\
    P11 & Consulting & Lead Data Scientist (DS) & Revenue maximization strategy prediction \\
    P12 & Internet Services & Head of ML (DS) & Oversees more than 50 models \\
    P13 & Transportation & Senior Software Engineer (DS)  & Place recommendation, partners acquisition \\
    P14 & Social Media & ML Engineering Manager (DS) & Advertisement ranking \\
    P15 & Transportation S/W & Senior ML Research Scientist (DS) & Claim handling, driver voice assistance \\
    P16 & Healthcare & Medical Doctor (DS) & Mortality, deterioration, and radiology \\
    P17 & Manufacturing & Senior Researh Scientist (DS) & Oil pump/power grid operation \\
    P18 & Software/Research & CTO (SE) & Tooling (image-based model interpretation) \\
    P19 & Finance & Senior Analytics Expert (PM) & Credit score prediction \\
    P20 & Healthcare & CTO, Head of Data Science (DS) & Lung cancer operation \\
    P21 & Software & Principal Design Manager (UX) & Best images selection, transcription \\
    P22 & Healthcare & Senior Data Scientist (DS) & Care management/resource utilization \\
  \bottomrule
  \end{tabularx}
  \vspace{2mm}
  \caption{Interviewees demographics, from the left: (1) Participant ID (PID), (2) Domain of their companies, (3) their job title and main role (DS for Data Scientist, SE for Software Engineer, DE for Data Engineer, PM for Product Manager, and UX for UX Researcher), (4) Model problems participants mentioned in the interview.}~
  \label{tab:table1}
\end{table}

Table~\ref{tab:table1} shows details of the 22 individuals who participated in our interviews. We had 22 individuals from 20 different companies (6 females and 16 males). Our participants span a wide variety of domains including banking, finance, healthcare, software companies, social media, transportation, consulting, manufacturing, internet services, and transportation. Among our participants, 17 described themselves as data scientists or machine learning engineers, 2 identified themselves as software engineers whose goal is to build infrastructure related to model interpretability (model data pipelines, model interpretability tools), 2 identified themselves as product managers. One identified himself as a UX researcher who has been collaborating with ML engineers to build models used in various product applications, such as personal photograph ranking, face recognition, and voice transcription. All of them worked in teams of data scientists and engineers who need to communicate with product managers, customers, and other stakeholders.

\subsection{Interview Guide}
We used a semi-structured interview approach. To prepare for the interviews, we developed a slide deck with a script and a set of questions organized around the following main themes: (1) participant's background and role; (2) type of projects developed in the organization; (3) experience with interpretability in their work (when, why, how interpretability plays a role, and who else it involves); (4) challenges commonly faced in the organization and desirable future developments. We shared the slide deck with the participant and used it to direct the conversation during the call. We provide the interview protocols in the supplementary material.

Before starting an interview, we obtained permission from our interviewee to audio-record the interviews. We also noted that we would share the outcomes of the study with the participant in the event they wished to provide further comments. In total, we collected 19 hours and 10 minutes of audio recorded interviews from 22 interview sessions. Each interview lasted an average of 52 minutes, normally lasting just a few minutes longer or shorter.

\subsection{Qualitative Coding Process}
All the audio recordings were transcribed by professional transcribers. The first and third authors worked together to perform an initial qualitative analysis, with the second author corroborating results and contributing further insights from an independent pass after the initial round of analysis. The analyses we present represent the perspectives of our participants systematically curated in an interpretive framework which was developed through discussions among the authors.

We used an iterative \textit{qualitative coding} process~\cite{saldana2015coding} characterized by alternate phases of coding (to tag specific text segments with codes), analytic memo writing (to collect ideas and insights) and diagramming (to build themes and categories from the codes). More specifically, we moved through the following steps. In the beginning, the first and third authors each created, using a subset of the interviews, initial sets of codes and memos (i.e., open coding, pre-coding~\cite{layder1998sociological}). Then, we shared our work and compared our codes to find commonalities and discrepancies and to agree on a common structure. At this stage, a small set of anchoring concepts emerged from open coding - roles, processes, goals, and strategies - which were influenced by our goal of understanding how interpretability manifests in real workplaces. We also explicitly noted descriptions of where roles interacted around interpretability, conceptual definitions and tensions perceived around interpretability, and challenges relating to achieving interpretability in the organization. Equipped with this knowledge, we went through the second round of coding, this time including the rest of the interviews and keeping track of insightful quotes. At this stage, we met again to share our work and to come up with a newly refined common structure emerging from our coding activity, with the remaining author joining in the discussion after an independent code pass. Finally, we reviewed together all our coded text snippets and memos to tweak and enrich our structure with relevant details. The final structure is reflected in the organization of this paper and comprises: roles, processes, goals, and a set of cross-cutting interpretability themes.

\section{Results}
We organize our results around the anchoring concepts we identified in our qualitative coding: (1) Interpretability \textit{Roles}, describing `Who' is involved, (2) \textit{Stages}, describing `What' activities take place and `When' in planning, building and deploying, and managing ML models, and (3) \textit{Goals} describing `Why' interpretability is perceived as being sought. We then describe themes that emerged from the intersection of these anchors which extend and problematize prior characterizations of interpretability.

\subsection{Interpretability Roles, Stages, Goals}

\subsubsection{Who?: Interpretability Roles}
We organize the stakeholder roles described by our participants around three main categories: Model Builders, Model Breakers, and Model Consumers.

\vspace{2mm}

\textbf{R1. Model Builders}: Model builders are individuals responsible for designing, developing and testing models as well as for integrating them in the data infrastructure of the organization. The most common job title of professionals found in this category are \textit{Data Scientists} (or often called ML Engineers) and \textit{Data Engineers}, with the former typically responsible for model development and validation and the latter responsible for integrating the model in a larger data infrastructure. 

\vspace{2mm}

\textbf{R2. Model Breakers}: Model breakers are people who have the domain knowledge to verify that models meet desired goals and behave as expected, but may not necessarily have professional level of knowledge about ML. They work with model builders to give them feedback about things that need to be improved in building their model. Model breakers contribute different perspectives to help identify and fix potential issues with the model, hence the name ``model breakers''. Within the category of \textit{model breakers} we identified three main sub-classes of breakers.  \textit{Domain Experts}, who are individuals with a deep knowledge of the real-world phenomena the model is supposed to capture (e.g., physicians in the development of a healthcare product, business analysts working for building credit evaluation models). \textit{Product managers}, who are responsible for the development of the actual product the organization needs. They communicate requirements and business goals and verify that they are met. \textit{Auditors}, who examine models from the legal perspective. Their goal is to make sure the model satisfies legal requirements from the compliance standpoint.

\vspace{2mm}

\textbf{R3. Model Consumers}: Model consumers are the intended end-users of the information and decisions produced by the models. This group includes a large variety of professionals who aim at making high-stakes decisions supported by ML models. Examples include physicians working with mortality and readmission models; bank representatives who use loan approval models for handling their customer's loan request and explaining outcomes, engineers predicting failures with expensive and mission critical machinery, and biologists testing potentially adverse effects of new drugs. Many of the professionals we interviewed explained that often the actual end-users of a model are other individuals employed in the same company they work for.

One important aspect to keep in mind regarding these roles is that they do not necessarily define a single individual. At any given time, the same individual can cover more than one of the roles we have outlined above.

\subsubsection{What? When?: Interpretability Stages}
Our participants described in detail how interpretability plays a role at many different stages of the model and product development process.

\vspace{2mm}

\textbf{S1. Ideation and Conceptualization Stage}: Before developing an actual model, model developers spend considerable time exploring the problem space and conceptualizing alternative strategies for model building. We found that interpretability plays a major role already at this stage of the process, even before an actual model is developed. For example, P11, referring to feature engineering, remarked: \textit{``... this is the first step toward making interpretable models, even though we don't have any model yet.''}. In particular, we found several data scientists complement feature engineering considerations pertaining exclusively to model performance with considerations on how feature engineering (e.g., adding, dropping out, re-scaling, combining features) would affect interpretability.

In performing feature engineering, we found model builders often involve model breakers to make sure the choices they make are not going to affect negatively interpretability needs that will surface later on in the process. Builders may thus involve domain experts to help them design feature sets that ``make sense'' from the point of view of their domain knowledge. For instance, participants in finance (P3, P5, P6, P9) commented about the importance of having the perspective of business analysts early on in defining features for credit evaluation models. P9 mentioned: \textit{``So, typically I would sit down at the very beginning of that process with the client, that business stakeholder, and essentially ask them to describe to me, in their words, what that sort of data generating process, how they would expect that to look. So, not obviously in terms of machine learning, but in terms of sort of their sense of what drove the outcome. What sort of variables would be important for someone to be interested in this particular product or whatever the case may be. [...] I would spend time with those teams going through essentially data dictionaries and then getting their point of view on how applicable or not applicable variables were for each of these objectives.''}

Similarly, we found that some model builders and auditors work together (especially in heavily regulated industries) in the early stages of the process to write a \textit{``model white paper''} that explains the meaning of each feature, how features are engineered, and why using the set of feature does not pose risk of a legal violation. P5, who works as a data engineer in a major bank explained that designing features together with the compliance team makes it easier to gain their trust later on in the verification and validation stage: \textit{``You have to have an application. This is why we (data scientists) think we need the model, these are who we think it will impact, ... and to convince the model risk officer that this is why we're doing it, this is the reason, this is why we know or believe that this will not cause any biases to occur in our model. Because yeah, you can't have a model that has lots of biases and that's a lot of legal problems down the road in bank industry.''}

\vspace{2mm}

\textbf{S2. Building and Validation Stage}: In this stage, model builders develop model prototypes and investigate them using various interpretability tools. An interesting finding from our analysis is that experts typically approach the problem using a combination of three main ``interpretability lenses'',  namely: a \textit{focus on instances}, for investigating individual or small groups of cases, a \textit{focus on features}, for understanding the impact groups of features have in making predictions and a \textit{focus on models} for investigating strengths and weaknesses of different models. In the following, we provide more details about each of these ``interpretability lenses''.

\textit{Focus on Instances}. A common validation strategy in a focus on instances is to build ``test cases'', that is, specifying and examining sets of cases for which the models should behave as expected. Determining these cases is typically done in collaboration with model breakers who may have more complete knowledge of the domain and business needs. Complementing this activity, builders are often explicitly seeking for “edge cases” (also often called ``anomalies'' or ``corner cases''), that is, cases for which the model is particularly uncertain or may behave unexpectedly. Even if limited in scope and completeness, reasoning around cases is the way builders often go about interpreting and testing a black-box model. 


A relevant issue regarding the use of test cases for validation is the need to understand how and why a model produces a given output for a specific case of interest; especially when the model does not behave as expected. This is where recently developed attribution/explanation methods such as  LIME~\cite{ribeiro2016should} and SHAP~\cite{Lundberg2017SHAP} can potentially play a role. Many of our experts mentioned experimenting with such methods as a way to develop intuitions about how a model makes a specific decision. However, the feedback we received has often been one of general uneasiness in using these methods; mostly due to their instability and limited transparency. For instance, P9 remarked: \textit{``LIME, to me, was just highly volatile, depending on the seed, depending on the features that you were using, I would just end up getting very different results, each run. ... [but] you get a consistent result with SHAP. I'm not saying that's good enough. ... Something with even SHAP is that ... it's an attempt to make an estimate. So, it's making a marginal claim in most cases.''}





\textit{Focus on Features}. A focus on features is also commonly described by our participants as an interpretability lens used to better understand a model. Builders often cite testing the ``plausibility'' and ``suitability'' of a model by looking at what features drive most of the decisions (typically by ranking them in terms of importance). Interestingly, when analyzing feature importance inspecting less important features can be as important (if not more) as inspecting the more important ones. For instance, P2 mentioned: \textit{``I also find that I've had a lot of success with feature importance because usually, the most important features are rather banal things. ... The relationship between the strongest feature and prediction outcomes pretty much explain well about the model, you know that that's a very obvious one. And then you get down to like, the fifth, sixth important features, and that's the area when you start to get into things that are novel and maybe that the business could influence.''} Information about feature importance is commonly shared with other stakeholders and helps them to reason about how the model behaves.



\textit{Focus on Models}. Finally, one of the most recurring tasks that model builders mentioned as crucial and frequent in this stage, is comparing multiple models. Many participants mentioned that interpretability is not confined to understanding one single model, but more about an evolving set of models they progressively build, compare and refine (often including legacy models versus new models). Alternative models could be built to, e.g., test different sets of hyper-parameters, alternative sets of features, different data sets or learning algorithms. Even though many participants mentioned model comparison as essential in carrying out their model interpretability tasks, it seems there is no established methodology or common set of tools that can be used for this purpose.



Our participants' responses suggested two overarching interpretability needs that emerge in this phase. First, the need for model builders to \textit{gain confidence} in the reliability and validity of models they build. Second, the need to \textit{obtain trust} from other stakeholders. In relation to these needs, model builders seem to have a general uneasiness with ``black-box'' models in which the logic used to make decisions is inscrutable or too complex to guarantee a high level of trust. This problem is particularly acute in high-stakes environments where even a single mistake can have very dramatic or costly consequences. P22, a senior data scientist working on a model for Intensive Care Units (ICU) remarked: \textit{``So for sure ... I don't think we can get away with providing a black-box model. I just don't think that's going to work as much as people say it ... if you don't explain it to them (model consumers), I don't think they will ever want to use it.''} When discussing what is lacking with current interpretability tools for examining neural network-based models, the majority of our participants desired better tools to help them understand the mechanism by which a model makes predictions; in particular regarding root cause analysis (P1), identification of decision boundaries (P3), and identification of a global structure to describe how a model works (P13).


As mentioned above, model builders have, in addition to model analysis and validation needs, a strong need to communicate to other stakeholders how models behave, how much they can trust them, and under what conditions they may fail. Many mentioned that the quality of communication matters for encouraging insights that they can use for improving models and to obtain organizational trust. Communication between different stakeholders is often iterative and may motivate further modeling to capture insights about how a model is doing. P16 described this need as follows: \textit{``We have meetings about every three weeks where we give feedback about the models we're building. Then they come back to us and say, `That's great, or this can be better, it could be worse.' Or something like that. We just kind of do that over, and over, and over again.''}

One of the strongest needs model builders described involves devising methods and tools to help them interact and communicate with model breakers. Involving breakers in the validation process using language and representations they can understand is one of the main challenges surfaced from our interviews. P18 mentioned: \textit{``I think there should be ways for you to communicate this information outside of your data science group, because especially in AI applications, the stakeholders that are involved are a fairly diverse group.''} Many of our participants mentioned a ``curse of knowledge'' (P7, P10); model builders often had difficulties in identifying what other stakeholders don't know. Even if they did grasp stakeholders' knowledge level, determining how to deliver the insights seemed difficult, even with the help of visualizations.

\vspace{2mm}

\textbf{S3. Deployment, Maintenance and Use Stage}:
Once a model is fixed, the next step is to deploy it by integrating it to a data infrastructure with real data streams. We found that interpretability issues in this stage are not circumscribed to a single model but to a whole infrastructure in which the model will take place. In this sense, interpretability issues are not limited to understanding how a model behaves but also how a whole infrastructure built around it behaves. This quote we collected from P4 exemplifies this problem: \textit{``[Before integration] I'm like talking to the data scientist and [I am] like: 'please look over all the parameters', and she's like: 'oh, just use the model with the parameters you used last time' ... So that's more like a data engineering bug, right? ... I don't know if it's something in the model that's incorrect, that's causing [the problem]. So that's the grueling part. Figuring out which side has the problem and then getting the right people to fix it.''} Several participants' comments hint at a tension between considering interpretability as the need to understand how a single model works versus how a whole infrastructure based on one or multiple models works. When models are considered in the context of the ecosystem they live in, participants suggested it is important to realize that interpretability challenges may stem from problems with the whole software system rather than the single model the system is based upon.

When asked to describe when interpretability challenges arise in the life-cycle of a ML system, many participants mentioned issues arise \textit{after} a model has been put in production. Relevant problems are how to check instances/prediction patterns that cannot flexibly reflect changes of the ``state of the world'', and when models make ``unacceptable errors'' (P10) in supporting high-stake decision-making situations. Such cases are to be detected by the team who is responsible for monitoring the behavior of the model. Once an interpretability error is identified, model builders try to understand why the model does not behave as expected and how to intervene on the system to fix it. As explained above, this type of ``root cause analysis'' can be particularly challenging. As P5, a data engineer, remarked: \textit{``Sometimes I feel like I'm doing some worrying for the data scientist [for detecting errors caused by some change from the data stream]. I think for like the two-year historical run, that's something that computationally takes a lot of time to do.''}. P2, P3, P7, P10, P13, and P15 remarked that such a case may lead to inspecting elements of the system that go beyond the specifics of a single model. Even though, it may take ``multiple days'' to just see the results (P5) due to the sheer size of the data.

Similarly, data science teams are often tasked with building a better version of a ML model once it has been decided a new version may improve performance or address new needs. In this case, a common interpretability challenge is also closely related to model comparison. Comparison between a new version of a model with an older one is essential but also currently challenging due to insufficient tooling support (P3). In general, to support their analysis of how an update affects interpretability, model builders intentionally keep model inputs constant. P20 remarked: \textit{``[When improving the model] the datasets we use are generally the same. We compare models using different parameters for tuning hyper-parameters or using model bases.''}

Since the stage after deployment is concerned with interfacing with end-users, this is where participants described \textit{explanations} or \textit{justifications} for model decisions being most often requested. Interpretability is a major concern when considering the interface a ML system provides to the actual model consumers. There were two main situations where explanation become crucial.

The first one is when model predictions are used under high stakes conditions. For instance, physicians using prediction models as an aid to medical decisions involving patients, need to understand how a model ``reasons'' in order to integrate the knowledge derived from the model with their own mental representation of the problem. P20 noted, referring to a model that helps physicians to determine procedures for patients with lung cancer, that physicians \textit{``almost always seek for explaining the prediction''}. P22 added that even explaining a prediction may not be enough and that physicians would rather have information about what aspects of a patient could be manipulated. Explanations for model predictions can play multiple roles: they can be used as ``evidence'' (P16, P18, P20) to corroborate a decision, when ML model and user's mental model agree, or they can even generate new insights when they disagree. When a mismatch between the two models occurs, without explanation, it can also lead to frustration and progressive mistrust toward the ML system. P18 mentioned: \textit{What happens when the human disagrees with the machine? That's the place where interpretability is also very important. The whole premise of intelligent decision making is you [an ML model] want to also help humans to be more intelligent. But if you're all of a sudden making predictions that are better than the human predictions, but you can't really explain them, you can never exchange this knowledge with your humans that are involved in this decision making process ... They (humans) can't learn, they can't improve their own internal predictors.}

The second one is when a justification is needed for a model's decisions that go against the desired goals of a model consumer. A typical case is models used in banks to support representatives meeting with their clients. P6, the AI head at a bank responsible for business credit evaluation, mentioned two main situations in which explanations are crucial in the organization. First, when a customer's loan is denied, it is very important for the representative to be able to provide a justification (the ``Right to explanation'' is part of the legal code of many countries around the world). Second, representatives have a strong interest in granting more loans, therefore they are always seeking explanations they can act on, for example, those that suggest to them how to turn declined loans into accepted ones. P9 noted: \textit{``The problem happens when the loan is denied. When it is denied, often the case, sales reps are even more curious to know why it was denied [than customers]. If the reason cannot be explained, now that can be a business problem.''} Hence, interpretability can play a role within an organization in mediating information among multiple actors to support problem-solving.

In summary, the stages we identified suggest that interpretability work takes different forms throughout the model lifespan. In S1, we found that interpretability is mostly focused around selecting and engineering features and bringing people with different roles together to define goals and to build trust. In S2, model interpretability work is aimed at tuning models (often by comparing multiple model versions), surfacing possible edge cases, seeking plausible and interpretable feature sets, and in particular, helping model builders communicate with model breakers and model consumers. In S3, model interpretability is about understanding the root cause of problems surfaced by monitoring systems and customers, comparing new and old models, and providing explanation interfaces to end-user for support to decision making and insight generation.

\vspace{2mm} Above all, a common feature of all of these stages is \textit{communication}. Interpretability efforts are often prompted by, and happen through, communication between several actors. Often interpretability work is not only about the use of specific techniques and tools, but also about the exchanging of information about a model's goals, needs, and workings among different people in different roles within an organization.

\subsubsection{Why?: Interpretability Goals}

The previous section touched upon specific goals various stakeholders may have at different stages of model development. In this section, we summarize the set of goals we encountered in more detail to more explicitly connect goals to roles and processes. We identified three broad classes of interpretability goals.

\vspace{2mm}

\textbf{G1. Interpretability for model validation and improvement}: Being able to identify issues with a model and devise ways to fix it is one of the most important activities for model development. As described above, model builders and breakers are mostly responsible for this goal, though they frequently need input from other stakeholder roles to carry out validation and improvement tasks. For instance, auditors and legal teams play a prominent role since their primary goal is to make sure the model satisfies a stringent set of legal requirements. Many participants mentioned that relying on aggregate statistics is not enough and cannot be trusted as the sole way to validate a model; hence the need for interpretability methods and tools. 

In addition, models can learn non-meaningful relationships with high accuracy between input and output~\cite{caruana2015intelligible}, making the involvement of domain experts necessary for proper interpretation of results. While there are methods available for identifying model mistakes and edge cases, identifying irrelevant correlations and false implications of causation may require the involvement of domain experts who understand the phenomenon of interest in detail. This process is characterized by contrasting model behavior with the stakeholders' perceptions of the actual meaning of the data and model predictions. P16, a physician who works with a mortality model, mentioned that their debugging does not rely on merely seeking high accuracy, but also on preventing \textit{``catastrophic misclassification''} which can cause serious consequences (e.g., failing at treating critical patients in ICU). Again, domain expertise figures prominently in identifying what sorts of misclassifications qualify as catastrophic.

Some participants mentioned auditing as one of the important reasons they need interpretability. From a legal standpoint, transparency is needed to make sure models are not violating legal regulations such as the ``Fair trading act'' (P3), or General Data Protection Regulation which aims at preventing discrimination. P18 and P22 mentioned that when a model makes a mistake it is crucial to explain why a given decision was made. Overall, a very large proportion of our participants mentioned compliance as one of the main reasons why interpretability is needed.

\vspace{2mm}

\textbf{G2. Interpretability for Decision Making and Knowledge Discovery}: An obvious area where interpretability is crucial is when models are used by consumers to aid decision-making. Interpretability in these cases becomes a primary feature of the developed product: decision-makers just can't trust or use predictions that do not provide a rationale for their recommendations. Similarly, explanations need to match the expert's needs and the constraints of the domain.

In service of supporting a decision-making support goal, participants often described interacting with domain experts or other end-users of the models to understand what aspects of a model help them to make their decisions. And one of the recurring themes was actionability. P22 mentioned: \textit{``The doctor said, these are the five things you always look at. Okay, now the prediction model has to accompany with this five things that they always look at ... You can't just say I'm going to rank 200 features and then that the doctor goes seek it out ... if I just take the top 10 but not actionable [features] ... that's not very useful either, right?''} For P22, understanding what features of a decision a model user could act on took multiple discussions before they could curate the right set of information in the interface, those features that the decision-maker believed were relevant and actionable.

Some participants described the potential for a tool to explain its decisions as a catalyst for learning or knowledge discovery. 
For example, multiple participants emphasized that in a decision support context, model interpretability becomes critical when a decision-maker and a model disagree and that learning is an important result of interpretability in these situations. Every participant in healthcare and finance mentioned that without having an explanation of why a model makes a prediction there is no way a human can reconsider a decision or learn from a model. Several participants alluded to implicit goals that they had as model developers of enabling decision-makers to ultimately make better decisions by learning from the model.


Interestingly, we heard from several participants about situations in which their models are not necessarily built for decision-making or automation, but primarily as a vehicle to generate insights about the phenomena described by the data. For example, P11 mentioned a model they built to understand KPI (Key Performance Indicators) of surgeons' performance in the future. The goal was to understand which features drive performance rather than to predict future performance. 

P8 developed a user retention model that shows how likely a user will return to the service depending on which system feature(s) she used in the past: \textit{``the goal here is not to improve the prediction accuracy as much as possible. Although we want them all to be accurate, to some extent, we are more interested in learning is what the internals of the model tell us about the factors that correlate with user retention. So, that's somewhat different from maybe some other application in machine learning where the accuracy is our primary importance.''}

P17 shared a story about building a model to detect anomalous oil pump operations: \textit{``Stopping the pump will cost money. If you stop, then your manager would ask why did you stop it. Without understanding why, you could prepare some documentation but perhaps you may not be able to make the best argument. But if the model can facilitate your reasoning about why, which factor made the pump failed, that could help you to prevent the same case happening in the future. That's what we call knowledge discovery.}'' P8 similarly described their need to understand customer retention in a social media application. The model they built had the main purpose of generating insights about what makes customers come back to their product rather than to make predictions: \textit{``the goal here is not to improve the [model] as much as possible ...  [but] what the model [can] tell us about the factors that correlate with user attention. So, that's sort of somewhat different from maybe some other application ... where the accuracy is our primary importance.''}

\vspace{2mm}

\textbf{G3. Interpretability to Gain Confidence and Obtain Trust}: Virtually everyone in our interviews commented on the need to gain trust in a model. Many participants stated that debugging is essential for them to gain confidence in the fact that their model is working properly. This individual-level trust is typically the focus in ML interpretability research to date. Even in cases where transparency is not required by customers or stakeholders, our data scientists said that thoroughly investigating their models in order to build understanding and trust as part of necessary \textit{``due diligence (P2)''}. However, many of our participants also emphasized that persuading others that a model can be fully trusted often takes precedence in their interpretability work. They described how a lack of trust in the enterprise of model building, or the work of the team developing those models, could prevent progress in the use of ML in the organization. Hence, model builders not only need to gain trust in their models and procedures but also may rely on interpretability to persuade others in different roles in the organization that their work provides value. P18 summarized the need many of our participants felt about how collaborating closely with other stakeholders can increase organizational trust: \textit{``obviously you need to be very close to the people who are building the models, and then very close to the people who are consuming the model output, or who are considered experts in the task that you're trying to model ... This aspect is very important because they bring a perspective of what makes software that we're using right now reliable, elegant, available, and scalable.''}

Being able to concisely and effectively explain how a model works and why it should be trusted is not easy, especially because some of the stakeholders may need to understand this at a high level of abstraction and with little technical detail. Several participants mentioned that visualization methods can play a major role in the communication stage of interpretability.

The high priority on building trust as a modeling goal for model developers means that they perceive the need to build transparent models or to use model extraction methods to explain to others how a given model works. Many of our participants described this need as one of the major ``pain points'' of model interpretability. P11, one of our participants working in banking, mentioned that many colleagues in his organization are still \textit{``scared''} to adopt black-box models for this reason.

\subsection{Themes: Characterizing Interpretability Work}
Research on interpretability aimed at producing new tools or insights has primarily focused on the interaction between an individual and a model. Interpretability is described as a property of the model class in general based on its relation to human expectations (e.g.,~\cite{lage2018human}) or constraints like monotonicity~\cite{gupta2016monotonic} or additivity~\cite{caruana2015intelligible}, or
human-driven constraints that come from domain knowledge or the presentation of a specific model's results that enables a human to understand why a label was applied to an instance (e.g., LIME~\cite{ribeiro2016should} or SHAP \cite{Lundberg2017SHAP}). 

While many of our participants described understandings of interpretability that corroborate its importance to building an individual's trust in a model and its role in processes like model debugging, several prominent themes in our results stand in contrast to prior framings of interpretability. We summarize four themes that emerged from our results which extend and may problematize existing characterizations.

\subsubsection{Interpretability is Cooperative}
A theme that recurred again and again in participants' descriptions of how they conceived of and practiced interpretability in their organization was one of collaboration and coordinating of values and knowledge between stakeholder roles. This theme was firstly exemplified by mentions of discussion with other stakeholder groups, such as domain experts, by more than half of our participants, who described these discussions occurring frequently during ideation and model building and validation, but also after deployment. Many participants explicitly remarked on how collaboration around interpretability was important for improving their business reasoning and convincing them that the models they built brought value. As P9 remarked: \textit{``including domain experts at the very beginning of the process is absolutely vital from an efficiency standpoint. And ultimately from the standpoint of building a good data product.''}. P22 described how inducing different roles' perspectives can produce clear modeling implications for their team which would be difficult to arrive at in other ways: \textit{``And sometimes we realize that certain features are maybe not properly coded. Sometimes certain features are overwhelmingly biasing ... sometimes it's a new discovery like, oh, we never thought of that.''}

Interpretability's role in building and maintaining trust between people in an organization, as mentioned by many of our participants, was another clear way in which its social nature was evident. Trust and building of common ground were perceived by some participants as resulting from communication affordances of interpretability: \textit{``We can easily use the interpretability using the visualization to convey this information to data scientists or even to the operation people. And they can fully use this information to convey the message to the rest of the operation. Interpretability is about communication between people, not necessarily just about person and model'' (P13)}. Similarly, P2 described how \textit{``at the end of the day, people respond to use cases and anecdotes way better than they respond to math, so I find that the proof and the need for interpretability is something I kind of have to put on myself, and when I'm communicating things to a client, it's usually just to explain the operation of the system and then giving some answer.''} 

Other participants explicitly commented on how trust gained through interpretability benefited themselves and/or their teams. When public relations, tech transfer, or other teams further from model development gained trust through interpretability, limitations or constraints on models tended to be more easily accepted with a model (P1). P9, who had spent nearly a decade working on customer-focused modeling in several large companies, described how internal trust was \textit{``always kind of the biggest deal in sort of that first thing that was always a necessity.''} The internal trust helped free his team up from constraints on the types of models they could use, or level of monitoring required, such that model interpretability was seen as a means to obtaining a certain status: \textit{``it became a means to allow that sort of flexibility, by gaining the trust of internal audit, be that legal or compliance, fair lending, there's a lot of groups that oversee this ... it took time to sort of build equity with those folks to get them to trust the work that I was doing and also to have trust that they would be able to communicate it. So giving them the tools, giving them the plots, giving them the information that they needed to be able to then carry that up the chain, was certainly a work in progress, but as we kind of get deeper into that, everyone became more comfortable.''}

A few participants implied that sometimes interpretability's importance was not that it could explain any given decision, but that the act of including it alone signaled a ``due diligence'' that other stakeholders or end-users found comforting. P21 implied how in his organization, the transfer of trust to people building models that interpretability facilitated could at times approach a harmful social influence: \textit{``They are saying, `I can build this', and `We can do that' and `We can go forward.' So if enough people around you are all rushing forward and they're saying `Hey we can all trust this AI thing' ... we are evolutionarily not inclined to disagree with them. We are instead inclined, socially, to say 'Great, it's working there, let's build on top of it.' So we get layered assumptions on top of assumptions, until the point where we end up with machine learning models that are trying to do things like understand how people feel.''}

\subsubsection{Interpretability is Process}
Existing characterizations of interpretability, such as by measuring mismatch between a human mental model and an ML model rarely comment on the potential for interpretability to occur in time. That our participants described interpretability work occurring throughout the model lifespan suggests that interpretability is a persistent concern. While interpretability might manifest differently across stages (e.g., as feature selection early in the pipeline versus identification of a model's consequences to existing decision pipelines after deployment), participants seemed to naturally associate it with a diverse set of practices and intents that might evolve or co-occur but which all contributed to their understanding of what interpretability means in their organization. From P16's reference to meetings with stakeholders as something his team did \textit{``over, and over, and over again,''} to other participants descriptions of how monitoring a model's use frequently reopened interpretability issues, many of our participants seemed to perceive interpretability more as a target for teams to organize work around than as a property that could be definitively achieved. As P17 described \textit{``talking with domain experts certainly is a challenge in this case and it's always iterative in industry. So it seems like you're training them and you're done. But it's also like improving and back to talking to the main experts again.''}

Several participants also described interpretability in light of its ability to foster ``dialogue'' between a human user and a model, which was perceived as necessary to maintain during the model lifespan for learning and continued use. For example, P21, a user experience researcher on an ML team, described how \textit{``useful interpretability isn't the ability to make the black box into a glass box, but rather the ability to perturb it, to ask it questions and say, `Hey, tell me what you do in this scenario,' like counterfactuals or, `Tell me what you've done in past scenarios, so I can interview you and learn about you.'''} Perceiving interpretability as enabling dialogue prioritizes implicit goals of learning and sense-making about a phenomenon that is being modeled, aligning with goals of knowledge creation. To P18, creating the ability for continuing information transfer between an ML model and a human was integral to the value of interpretability: \textit{``but if you can't really explain them, you can never exchange this knowledge with your humans that are involved in this decision-making process, and aside from having just psychological challenges to the humans who are involved in this because they can't understand why the machine is predicting, why the model is predicting better than they are. They can't learn, they can't improve their own internal predictors. I think that signifies that we're also not extracting as much value as we could be extracting from these systems that we're building.''} Multiple participants seemed interested in interpretability as a means of correcting poor understandings or decisions. We discuss the role of the model developer's sense of ground truth further below.

\subsubsection{Interpretability is Mental Model Comparison} 
The types of learning and sense-making that interpretability was perceived to bring about in an organization often occurred through comparisons between mental models held by different stakeholders.
This aspect contrasts somewhat with definitions of interpretability that frame it as determinable from the interaction between an individual mental model and ML model. While an ML model to human comparison was implicit in many of our participants' descriptions around interpretability, that the ML expert's own understanding of the problem had a mediating influence on how they pursued interpretability was apparent.

One obvious way in which this manifested was in participants' descriptions of how they deeply considered the mental models of other stakeholders, often domain experts and/or end-users, in their work. P21 described how his interpretability work hinged on his ability to internalize the mental models of his users: \textit{``We try to bring people more proximate, closer to the end-user ... instead of asking ourselves what we think would be awesome, and what level of interpretability seems necessary, let's ask the people who will be affected by it. What they need in order to have confidence.''} Multiple other participants described how their team devoted considerable effort toward understanding end-users mental models to inform their work.

For several other participants, reflecting on aspects of human decisions helped them refine and put in perspective interpretability goals. For example, P12 described frustration with unrealistic expectations on model interpretability: \textit{``So while people [...] are very skeptical about machine learning and how much we can explain things, we know what data the AI model has seen. We know what inputs it's acting on. When you ask a human to explain things, you don't even know what they're using. You don't know if they're taking to account the color of your skin. You have no idea. They don't know.''} Similarly, P21 described a frustration he felt when reflecting on human explanation abilities in light of expectations of ML models: \textit{``through the lens of machine learning, we often think that there should be a perfect mental model that a person can form with how a machine learning system operates... in a complex vision task, it's the same way it operates with humans. You can't ever ask another human being `Hey, why'd you do that? Why'd you interpret it that way?' I mean you can, but you will get a terrible response. That's what we're doing when we ask a complex machine learning system, 'Hey, why'd you do that? ... we refuse to use systems unless we believe we can fully interpret them. There's a problem there.''}

Additionally, the mismatches between an ML model and human mental model that our participants, as ML experts, described themselves being attuned to were often interpreted and addressed in light of their own intuitions about how the ML model was behaving. For example, P12 described how a critical insight for her team was a realization that shape constraints (manually added constraints about the relationship between a feature and the output value, like, e.g. monotonicity constraints) could be used to represent many human end-users hypotheses about how features in a model should behave. Once the team had a concise representation of the anticipated behavior, they could use these expectations to describe the behavior of the model in a way that provided guarantees that the end-users needed to gain trust: \textit{``I guarantee you that the GPA is never hurting and the LSAT scores [are] never hurting. And they can walk away feeling more like they can trust them all and more like they understood what the model is doing.''} Understanding human hypotheses in light of her expertise in ML was for P12 a large part of interpretability work. 

Similarly, the ML expert's mental model appeared to influence their choice of whether or not, and how, to revise an ML model based on an identified mismatch. Implicit in this comparison was an assumption on the part of the ML expert about what the ``correct'' response to a given instance was. This enabled the ML expert to make a judgment call about whether the mismatch was likely to be the result of a flawed human mental model or a problem with the ML. For example, some participants appeared to pay close attention to how models were being used, in order to monitor whether end-users were likely to be biased by a model or they could learn from it. P16 described how in their organization, \textit{``As far as the ones I know, it's always a human does a read first, then the computer gives a recommendation. That way the human is never biased by the machine.''} P16 continued \textit{``ultimately, my ideal situation is to understand when the model was right and be able to persuade a person that they should do what the model is telling them to do.''} This was also evident in some participants' comments about how certain mismatches were more critical to address. P18 described how: ``\textit{I tend to have an opinion that when the human is performing some kind of task, some kind of intelligent prediction task, they could be wrong a lot, but typically when they're wrong, they're not very wrong. They're not orders of magnitude wrong. They're within small range of deviation wrong. Because they have very deep understanding of this, and when they're wrong, they probably made one assumption wrong and that's what caused their forecast to be off. When you have an intelligence system that learned from data but has no real context of what it learned, and doesn't really understand what actually the forecast is, what demand is, how big demand could reasonably be, or how small could demand reasonably be, places where it could go wrong, it could go orders of magnitude wrong.}''

\subsubsection{Interpretability is Context-Dependent}
Related to our participant's dependence on understanding other human mental models was a sensitivity to context that often came through in their comments. As P12 described, her team's focus on identifying what hypotheses different groups of people had made interpretability \textit{``about finding the guarantees that different groups of users need.''}

Participants highlighted how interpretability solutions built by their teams were strongly shaped by the needs and use cases of the particular user groups for which they were intended. This included P22, who described the importance of doctors' perceptions of what features were \textit{actionable} or associated with treatments: \textit{``So, a lot of curation needs to happen with the domain expert in terms of what information is useful. Just providing, say, here are the top 10 [features], it just doesn't help at all ... The doctor said, these are the five things you always look at. Okay now the prediction model has to be accompanied by these five things they always look at.''} 

P16 confessed to initially being hesitant to provide interpretability for models he built in a healthcare context due to concerns that explanations would be used to infer causal relationships between outcomes and features. However, his fear was assuaged as he learned more about the population (of doctors) that he was designing for: \textit{``I was very hesitant to show explanations to people in healthcare, but now I really want to. I really want to show people explanations because at least the providers understand that there's no causality associated with a sign or a symptom. Those signs or symptoms are a consequence of a disease that's happening internally. If you fix that heart rate, it doesn't mean that the disease is fixed, right?''} Similarly, for P22, who also worked in healthcare, knowing her users, and in particular the skepticism they brought to their work, was critical to her motivation for providing interpretability: \textit{``all the predictive models that we want to have, have some form of explanation to go with it, that's our standard immediately'': ``if you are a nurse, if you are a doctor, you should always challenge every result that people gave you. Challenge it.''}

Some participants also implied that the impacts of interpretability work were intimately related to decision-making infrastructures that could be complex. P12 described: ``\textit{You have to figure out is that revenue gain coming in a way that's really truly beneficial for all of the stakeholders. And so there's a lot of analysis that will happen there and a lot of decision-makers and just I don't know, layers of decision-makers. These systems often are pretty interconnected and might also destabilize other things. So sometimes you might build a model and then maybe some other people are using it in some new way and you need to make sure you're not breaking anything from their point of view.}''

\section{Design Opportunities for Interpretability Challenges}

Our interviews surfaced a number of challenges faced by ML experts working toward interpretable models in industry. While an exhaustive listing of opportunities for technology to better support this work is beyond the scope of this work, we present a select set of challenges that we distilled from the issues raised by our participants. These challenges were chosen in part because they represent areas where existing research has not focused much attention, but where progress may have a strong impact on ML/HCI research and practice.

\subsection{Identifying, Representing, and Integrating Human Expectations}
The prominence of attempts to identify and handle edge cases in the work of our participants suggests that more systematic approaches to identifying these important ``anchors'' for interpretability work would increase interpretability-related task productivity. The edge case identification could happen in two ways, between a model and an individual, and a model and a group of people. 

One opportunity lies in eliciting and representing an individual's expectations in ways that are compatible with the model representation that ML experts are accustomed to, even if through proxies. During the interview, P12 discussed the importance of regulating the model's outcome when the shape does not align with her ``common-sense''. She described discovering shape constraints as a powerful representation that her team can use to easily regulate a model's behavior when the model doesn't align with the common-sense level of human expectations. However, which ``side'', the human or the model, is correct in the case of a mismatch may not always be obvious. P2 introduced his term ``superstition'' which refers to an inaccurate assumption that people can have: \textit{we knew what I called the superstition of the client that X was always true and this was important; so find out those internal expectations they had, most of which will be true and some of which will be folk wisdom that isn't factual. So it's [ML model] going to take those values that they're not expecting to look at.}

Resolving such tension between a human and a model is an active research topic in the CSCW and CHI communities \cite{zhang2019dissonance}. P12 mentioned tools that could help more quickly resolve discrepancies between a model's prediction and her expectations: \textit{``if your hypothesis is wrong and it really hurts the accuracy to make that constraint, then you have to ask, what's going on in there. I feel like we don't have very good debug tools necessarily to figure out why is the model not monotonic when I think it should be.''} We propose that interpretability research would benefit from interfaces and mechanisms to help a human (1) articulate their expectations around predictions in model interpretation tasks (2) efficiently recognize the gaps between their expectations and model representations, (3) gain context through further evidence available to the model that may help them assess the gap, and (4) enable them to debug/change model behavior based on their findings. Looking to research in belief elicitation and mental model discovery within psychology, economics, and related disciplines in designing such approaches may be informative (e.g.,~\cite{bostrom1992characterizing, ford1998expert, kim2017explaining, kim2017data, o2006uncertain}). Bayesian approaches to providing answers to users' queries based on conditional probabilities may be another promising approach~\cite{mansinghka2015bayesdb,saad2016detecting}.

Multiple participants described the discussions that they or their team members repeatedly engaged in with other stakeholders to identify such cases as difficult yet crucial. Our finding that data scientists often engage in such collaborative work aligns with recent findings \cite{zhang2020data}. Aligning schedules to make meetings happen was not always possible, leading at least one participant to feel regretful about missing the value that more extensive discussions with end-users could provide. Having access to other stakeholders' responses to model results was implied to be valuable during the stage of model development and verification. For example, P13 described how, \textit{``If I can proactively test or verify the interpreted results, it would be super helpful. So, now the models can give me some results. I can just look at the result and make the decision for myself. But I still have to define another set of rules, actions to verify if it actually works. If there's a way for us to easily, interactively, verify ...''} One approach that may be fruitful is to develop interactive applications that an end-user or domain expert can use to provide their (asynchronous) feedback more efficiently. While some research has demonstrated the efficacy of using large scale crowdsourcing platforms to identify errors made by predictive models~\cite{attenberg2011beat} or getting collective answers from a group of people~\cite{chung2019efficient}, we are not aware of research aimed at developing similar approaches that are better suited to smaller numbers of potentially expert users. Interactive approaches could provide both domain experts and ML experts an ability to engage asynchronously where needed, reserving valuable in-person discussion time for interrogating these instances or sharing knowledge about decision processes. Though many participants described challenges to gaining feedback, none mentioned having developed in-house tools to support this common task, nor having obtained tools from a third party.

\subsection{Communicating and Summarizing Model Behavior}
We observed several ways in which the communication that comprised interpretability work, between stakeholders within an organization or occurring as a form of ``dialogue'' between a human and a model, can often result in misinterpretation. Several participants remarked about the need for more accessible interpretability solutions. P20 said \textit{''I want interpretability for everyone,''} referring to solutions that would be appropriate for both domain experts and practitioners in the healthcare setting in which he worked. Similarly, P17 desired methods that came closer to \textit{``presenting the predictions in a colloquial way''} to facilitate easier interpretation on the part of users. 

The cooperative and social nature of much of the interpretability work our participants described also suggests that transferring knowledge in the form of model “hand-offs” happens frequently. In these situations, having representations, such as visualizations, that can concisely capture how those receiving a model might expect it to behave in relation to their mental models could be useful to provide the types of model “bug” discoveries that some participants described occurring unexpectedly after deployment. These representations could be developed along with the algorithms mentioned above for efficiently identifying important test or edge cases for summarizing a model’s interpretability status.

A related concern voiced by some participants involved finding ways to improve existing interpretability approaches to provide more direct and customized information on why a model disagreed with a human mental model. For P12, tools that could help more quickly resolve discrepancies between model behavior and representations of her own or her users’ expectations in the form of shape constraints seemed important: ``if your hypothesis is wrong and it really hurts the accuracy to make that constraint, then you have to ask, what’s going on in there. I feel like we don’t have very good debug tools necessarily to figure out why is the model not monotonic when I think it should be.''

One possibly fruitful research direction is research which focuses on hiding the complexity behind building an ML model and making model-building more accessible to non-specialists~\cite{wang2020humanAI, hong2019disseminating}. Another direction may be building more intelligible visualizations that can effectively present the behavior of a model to the general public. For instance, presenting uncertainty in model predictions using more intuitive visual abstractions~\cite{hullman2015hypothetical, kay2016ish} may be beneficial.

\subsection{Scalable and Integratable Interpretability Tools}
Many participants shared their concerns with integrating interpretability tools in their complex workflows and organizational infrastructure. Problems commonly cited are that: methods are not available because they were developed in academic settings, where creating robust and dependable software is not the norm; tools were not easy to integrate in existing platforms because they cannot be adapted to their specific environment; or simply that tools do not scale to the sheer size of their data sets. We introduce two emerging interpretability-related use cases: model comparison, and data preparation.

Participants pointed out that existing model interpretability approaches seem to overlook important, but perhaps less principled, aspects of approaches to comparing the outputs of multiple models to build a deeper understanding. We identified several different cases for model comparison: when comparing different parameters, selection of features, timestamps, and more. Though model comparison emerged as a ubiquitous task in model development, according to our results, existing interpretability tools have very limited support (if any) for model comparison (a notable exception being the Manifold tool~\cite{zhang2019manifold}).

Interpretability tools are also needed in the ideation stage of the model lifecycle. Better tools are needed before training to assist with ``data debugging'', to detect issues with \textit{data} before they are used for training, as well as for feature engineering and model design. Several participants suggested that there should be a better way for a model designer to test hypotheses about what effect different decisions on feature sets may have on model performance without necessarily building all these models entirely.

\subsection{Post-Deployment Support}
Multiple participants described their efforts to, and struggles with, connecting not only decision outcomes but the value of those outcomes to interpretability work. For example, for P16 \textit{``to have a framework for people to understand how the explanations either drive the decision making, improve decision making, or otherwise ... I have this model, but did it actually solve the business problem?''} 

In the post-training phase, better tools are needed to monitor models once they have been deployed. P16 also described how \textit{``The next step is when you're ... say you've communicated and everybody's happy with the model and you're ready to deploy it. Once you go out to meet with the engineers who are running the software stack, the first thing they're going to ask you is `How do I test and debug this? How do I continuously evaluate whether it's working?' And that's where your tools really really fall short.''}

Data visualization tools are one way to facilitate monitoring model behavior. Automated anomaly detection may be useful in addition, to identify and surface potentially troubling behaviors, given an appropriate specification of critical expectations that should hold such as one learned from domain experts. Multiple participants identified a need for better tools to support ``root cause analysis'': once issues with the model have been detected, it is crucial to have tools that can quickly point users to a potential source of the problem. As several participants alluded to, this is particularly hard in production because problems may stem from different modules of complex data architecture and not only from the model itself.

\section{Limitations}
While we devoted considerable effort to interviewing participants from a broad range of domains, our analysis misses other important domains, such as governance (e.g., predictive policing), agriculture (e.g., crop monitoring, animal health monitoring), and more. We focused on cases where model predictions are consumed by a human. We did not cover application areas where a model's predictions are used in fully automated and/or semi-automated settings (e.g., autonomous driving, or warehouse robots). Finally, many of our participants were data scientists with a technical background, therefore this work reflects their specific point of view, as interpreted by the authors through systematic qualitative analysis. Although we identified key stakeholder roles in model interpretability, we did not directly hear from every role that we identified as relevant to model interpretability. Future work might pursue ethnographic and other observational approaches to corroborate these results.
\section{Conclusion}
In this work, we empirically investigated industry practitioners' model interpretability practices, challenges, and needs. We describe roles, processes, goals, and strategies around interpretability practiced by ML practitioners working in a variety of domains. We draw out important contrasts between our results and the predominant framing of interpretability in existing research as a property of a model that can be defined solely in light of an individual versus a ML model's alignment. Our results lead us to characterize interpretability as inherently social and negotiated, aimed at fostering trust both in people and in models, context-dependent, and arising from careful comparisons of human mental models. Our identification of unaddressed technological needs among practitioners is intended to help researchers direct their attention to unaddressed, yet prevalent, challenges in industry.

\section{Acknowledgement}
We wish to express our deepest gratitude to our participants who were willing to share their experience and insights about their model interpretability practice and challenges. We also thank for DARPA Data-Driven Discovery of Models (D3M) Program for generously supporting this research. Any opinions, findings, and conclusions or recommendations expressed in this material are those of the authors and do not necessarily reflect the views of DARPA.

\bibliographystyle{ACM-Reference-Format}
\bibliography{99_REF}

\end{document}